\documentclass[printer]{aa}
\usepackage{graphicx}
\usepackage{txfonts}
\usepackage{amssymb}
\usepackage{booktabs,siunitx}
\usepackage{natbib}
\bibpunct{(}{)}{;}{a}{}{,} 

\usepackage{color}
\newcommand{\scb}{\textcolor{black}}

\def\be{\begin{equation}}
\def\ee{\end{equation}}
\def\bi{\begin{itemize}}
\def\ei{\end{itemize}}

\def\aap{A\&A}

\def\apjl{APJL}

\begin{document} 

\title{Probing scalar tensor theories for gravity in redshift space}
\author{Cristiano G. Sabiu\inst{1,2}, David F. Mota\inst{3}, Claudio Llinares\inst{4,3}, Changbom Park\inst{2}}

\institute{Korea Astronomy and Space Science Institute, Yuseong-gu, Daejeon, 305-348, Korea
\email{csabiu@kasi.re.kr}
\and
School of Physics, Korea Institute for Advanced Study, Dongdaemun-gu, Seoul 130-722, Korea
\and
Institute of theoretical astrophysics, University of Oslo, 0315 Oslo, Norway
\and
Institute for Computational Cosmology, Department of Physics, Durham University, Durham DH1 3LE, U.K.
}

\date{\today}

   \titlerunning{Probing scalar tensor theories for gravity in redshift space}
   \authorrunning{C. Sabiu et al.}

\abstract{We present measurements of the spatial clustering statistics in redshift space of various scalar field modified gravity simulations. We utilise the two-point and  three-point correlation functions to quantify the spatial distribution of dark matter halos within these simulations and thus discriminate between the models. We compare $\Lambda$CDM simulations to various modified gravity scenarios and find consistency with previous work in terms of two-point statistics in real and redshift space. However, using higher-order statistics such as the three-point correlation function in redshift space we find significant deviations from $\Lambda$CDM hinting that higher-order statistics may prove to be a useful tool in the hunt for deviations from General Relativity.}


\maketitle

\section{Introduction}
The $\Lambda$CDM cosmological model has been shown to reproduce many observational measurements, from the CMB at very early times to the late time clustering of galaxies. Despite the successes of the model, certain issues remain to be fully explained. These are either conceptual or arise from conflict with observational constraints.  One of the main problems is that the nature of the two main ingredients of the model (dark matter and dark energy) remains unknown.  Among the different solutions to the philosophical and quantitative inconsistencies associated with these components is  the idea of modifying the gravitational theory.  Several alternative theories exist \citep[][]{{2012PhR...513....1C}, {2012arXiv1206.1225A}}, all of which include extra degrees of freedom in the form of scalar, vector, or even tensor fields.  As General Relativity (GR) is proven to be valid to high accuracy on solar system scales \citep{2014LRR....17....4W}, any modification introduced must reduce to GR in these scales, which is done through screening mechanisms.  Within the context of scalar-tensor theories, there are three  such mechanisms based on conformal couplings, namely Vainshtein \citep{vain}, Symmetron \citep[][]{{2010PhRvL.104w1301H}}, and Chameleon \citep[][]{{cham}}.  In addition, \citet[][]{2012PhRvL.109x1102K} has recently proposed a mechanism   based on a disformal coupling.

Since all the above modified gravity models with  screening mechanisms seem to be viable, the question arises of  how to distinguish between the different models using cosmological observations. Since the effects of the forces associated with the extra degree of freedom emerge during the onset of non-linear structure formation, it is a promising tool that can be used to rule out some of the above models. The non-linear effects arising from these screening mechanisms have been studied using cosmological N-body {simulations }\citep{Oyaizu,Schmidt,Li, Lombriser, Boehmer, Zhao,2011PhRvL.107g1303Z,Barreira}. 

Second-order clustering statistics have been used to investigate differences between ${\Lambda}$CDM and modified gravity models, particularly those of  \cite{2011PhRvD..84j3521H,2011PhRvD..84l3524B,2012MNRAS.425.2128J} and \cite{2012JCAP...10..002B}.  Recently \citet{2015arXiv150701592S} have put observational bounds on $f_{R0}$ using SDSS data. \scb{Another novel technique for probing modified gravity involving second-order statistics was investigated by \citet{2015PhRvL.114y1101L} using the density-field clipping method \citep{2011PhRvL.107A1301S,2013PhRvD..88h3510S}. They found that constraints on $f(R)$ gravity can be tightened when clipping the density field with a high threshold since contributions of screened high-density regions to the matter power spectrum will be reduced and will boost the modified gravity effects.}

In this paper we  focus on higher-order clustering statistics in redshift space as a tool for differentiating modified gravity models (as alternatives to dark energy) from a fiducial GR model. Three-point  functions were studied in the linear regime for general scalar tensor theories   ranging from galileons to the most general Horndeski model \citep{2013JCAP...03..034B, 2014PhRvD..89j4007T, 2015JCAP...05..057B}.   The non-linear regime was studied by \cite{2011JCAP...11..019G} in the $f(R)$ case.  Here we repeat the calculations for the same $f(R)$ model, but extend this previous study in two ways.  Firstly, we run simulations with a code that includes an adaptive mesh refinement structure (AMR) giving a much more accurate description of the clustering, especially on small scales. Secondly, we focus our study not only on the real space correlation functions, but also on the redshift space functions, which are ultimately the ones that can be observed.  Furthermore, we also present predictions for the symmetron model in addition to $f(R)$.

The paper is structured as follows. In \S \ref{sims} we discuss the scalar field models investigated and briefly describe the numerical simulations  used. In \S\ref{stats} we lay out our framework for analysing the numerical simulations and discuss our specific implementation of the two- and three-point clustering statistics that we use. We present our clustering results for the various models in \S\ref{results}, and  conclude in \S\ref{conclusions}.

\section{ Models and simulations}\label{sims}
\subsection{ Gravitational models}
We focus our analysis on two specific scalar tensor models: the symmetron model and a particular case of $f(R)$ theories.  Both models include screening mechanisms, which reduce them to GR in  high-density regions and thus make them able to pass solar system tests.  Below we summarise the characteristics   of the models.  In both cases, we  work in the Einstein frame and so Einstein's equation will be unchanged.  The dominant dynamical effects will appear as a modification of the geodesics equation that is used to track the motion of the particles, which now will include a fifth force term.   

\subsubsection{ The symmetron model}

The symmetron model was originally discussed in \cite{2005PhRvD..72d3535P, 2008PhRvD..77d3524O} and \cite{2010PhRvL.104w1301H} as a standard scalar tensor model which has a particular coupling to matter.  The model includes a screening mechanism based on the restoration of a particular symmetry.  The cosmology of this model at the background and linear perturbation level has been studied in \cite{2011PhRvD..84j3521H} and \cite{2011PhRvD..84l3524B}. In the non-linear case, there are several results coming from quasi-static non-linear N-body cosmological simulations \citep{2012ApJ...748...61D, 2012JCAP...10..002B, 2014A&A...562A..78L}.  The effect of non-static terms in these simulations was presented in  \cite[][]{2013PhRvL.110p1101L, 2014PhRvD..89h4023L}.  

The action of the symmetron model is given by 
\be
S = \int \sqrt{-g} \left[ {\bf \frac{M_{Pl}^2}{2}} R - \frac{1}{2}\nabla^a\phi \nabla_a \phi - V(\phi)\right] d^4x + S_M(\tilde{g}_{ab}, \psi) \ ,
\label{symm-action}
\ee
where \scb{$R$ is the Ricci scalar}, the Einstein and Jordan frame metrics ($g_{ab}$ and $\tilde{g}_{ab}$) are conformally related
\be
\tilde{g}_{ab} = A^2(\phi) g_{ab},
\ee
\scb{and $S_M$ is the matter action which describes the evolution of the matter fields $\psi$.}
The potential and conformal factor that define the model are
\begin{eqnarray}
V(\phi) &=& -\frac{1}{2}\mu^2\phi^2 + \frac{1}{4}\lambda\phi^4 + V_0 \\
A(\phi) &=& 1 + \frac{1}{2}\left(\frac{\phi}{M}\right)^2, 
\end{eqnarray}
where $\mu$ and $M$ are mass scales, $\lambda$ is a dimensionless constant, and $V_0$ is tuned to match the observed cosmological constant.  The equation of motion for the scalar field that comes out from the action is then
\be
\nabla^a\nabla_a\phi = V_{,\phi} - A^3(\phi) A_{,\phi} \tilde{T}, 
\label{eq_motion_phi}
\ee
where 
\be
\tilde{T}_{ab} = -2\frac{1}{\sqrt{-\tilde{g}}} \frac{\delta L_M}{\delta\tilde{g}^{ab}}
\ee
is the Jordan frame energy momentum tensor \scb{(here, $L_M$ is the  Lagrangian matter)}. By fixing the metric to be a perturbed Friedmann-Robertson-Walker metric
\be
\label{metric}
ds^2 = -(1+2\Phi) dt^2 + a^2(1-2\Phi)(dx^2+dy^2+dz^2), 
\ee
\scb{where $\Phi$ is a scalar perturbation (i.e. the gravitational potential in a classical context)}, we can write the equation of motion of the scalar field in the  form
\be
\nabla^2\phi = \left(\frac{\rho}{M^2}-\mu^2 \right)\phi + \lambda \phi^3 = \frac{d}{d\phi}V_{eff}(\phi), 
\ee
where $\rho$ is the Jordan frame matter density and the effective potential is given by
\be
V_{eff}(\phi) = \frac{1}{2}\left(\frac{\rho}{M^2} - \mu^2\right)\phi^2 + \frac{1}{4}\lambda\phi^4 + V_0.
\label{def_effective_potential}
\ee
From this equation, it is possible to see that the expectation value of the scalar field vanishes at high matter densities. This sets the conformal factor $A$ to unity and thus decouples the scalar from the matter, producing the screening of the fifth force.

For numerical convenience, we work with a dimensionless scalar field $\chi \equiv \phi/\phi_0$, where $\phi_0$ is the expectation value for $\rho=0$:
\be
\phi_0 = \frac{\mu}{\sqrt{\lambda}}.
\ee
We also substitute the three free parameters $(M, \mu, \lambda)$ and use instead the range of the field that corresponds to $\rho=0$,
\be
\lambda_0 = \frac{1}{\sqrt{2}\mu} \ , 
\label{def_lambda0}
\ee
a dimensionless coupling constant,
\be
\beta_s = \frac{\phi_0 M_{pl}}{M^2} \ ,
\label{def_beta}
\ee
and the scale factor at the time of symmetry breaking,
\be
a_{SSB}^3 = \frac{\rho_0}{\rho_{SSB}} = \frac{\rho_0}{\mu^2 M^2} \ , 
\label{def_assb}
\ee
where $\rho_0$ is the background density at $z=0$.  Throughout the paper we  use either  $a_{SSB}$ or its associated redshift $z_{SSB}$. 
The equation for the dimensionless scalar field $\chi$ is then
\be
\nabla^2\chi = \frac{a^2}{2\lambda_0^2}\left[\left(\frac{\rho}{\rho_{SSB}} - 1\right)\chi + \chi^3 \right]. 
\label{eq_motion_chi}
\ee
In the Einstein frame, the effects of the scalar field on the matter distribution will be given by a modification of the geodesics equation, which takes the following form:
\be
\label{geo_code}
\ddot{\mathbf{x}} + 2 H \dot{\mathbf{x}} + 
    \frac{\nabla\Phi}{a^2} + 
    \frac{6\Omega_m H_0^2}{a^2} \frac{(\beta_s\lambda_0)^2}{a_{SSB}^3} \chi\nabla\chi = 0.
\ee
\scb{Here $H_0$ is the Hubble parameter at redshift $z=0$, $\Omega_m$ is the mean matter density at redshift $z=0$ normalised to the critical density,} and the dots represent derivatives with respect to Newtonian time defined by eq.~\ref{metric}.

\subsubsection{The $f(R)$ model}

Among the large number of $f(R)$ models that exist in the literature we choose the model presented in \cite{2007PhRvD..76f4004H}, which has attracted great interest in the context of dark energy models (see \citealt{2014AnP...526..259L} for a review on observational constraints on the model).  In the context of non-linear structure formation that we study here, there are several N-body codes capable of simulating this model \citep{{2008PhRvD..78l3523O}, {2012JCAP...01..051L},2013MNRAS.436..348P, 2014A&A...562A..78L}.  The validity of the quasi-static approximation assumed in these codes was studied in detail in the linear and non-linear regime in \cite{2013arXiv1310.3266N} and \cite{2015JCAP...02..034B}.

The action that defines the model is
\be
S = \int \sqrt{-g} \left[ \frac{R+f(R)}{16\pi G} + L_M \right] d^4 x, 
\ee
where the free function $f$ is chosen as
\be
f(R) = - m^2\frac{c_1(R/m^2)^n}{c_2(R/m^2)^n+1}, 
\ee
where $m^2 \equiv H_0^2\Omega_m$ and $c_1$, $c_2$, and $n$ are dimensionless model parameters. By requiring the model to give dark energy, it is possible to reduce the number of free parameters from three to two ($n$ and $f_{R0}$).  This requirement translates into
\be
\frac{c_1}{c_2} = \frac{6\Omega_\Lambda}{\Omega_m},
\ee
\scb{where  $\Omega_{\Lambda}$ is the density parameter associated with the cosmological constant}.  
Instead of using $c_1$ (or $c_2$) as the second free parameter, it is convenient to use
\be
f_{R0} = -n\frac{c_1}{c_2^2}\left(\frac{\Omega_m}{3(\Omega_m+4\Omega_\Lambda)}\right)^{n+1}, 
\ee
which relates to the range of fifth force in the cosmological background at redshift $z=0$ as  
\begin{eqnarray}
\lambda_\phi^0 = 3 \sqrt{\frac{(n+1)}{\Omega_m+4\Omega_\Lambda}}\sqrt{\frac{|f_{R0}|}{10^{-6}}}~~~ \mbox{Mpc}/h, 
\end{eqnarray}
\scb{where $\lambda_\phi^0$ is the range of the field, which is typically given in Mpc$/h$.}
It has been shown that the model can be translated into a scalar tensor model \cite[][]{brax}.  This can be done by applying a conformal transformation 
\be
\tilde{g}_{\mu\nu} = A^2(\phi)g_{\mu\nu}, 
\ee
where 
\be
A = \exp(-\beta_f\phi/M_{Pl}), 
\ee
with $\beta_f = \frac{1}{\sqrt{6}}$.  We also define 
\begin{eqnarray}
f_R = \frac{df}{dR} = e^{-\frac{2\beta_f\phi}{M_{\rm Pl}}}-1 \simeq -\frac{2\beta_f\phi}{M_{\rm Pl}}.  
\end{eqnarray}
This equation defines $R(\phi)$, which can be used to get the potential $V(\phi)$ in which the scalar field oscillates and that is given by
\begin{equation}
V(\phi) = \frac{M_{\rm Pl}^2(f_RR -f)}{2(1+f_R)^2}. 
\end{equation}
In the static limit, the scalar field $f_R$ fulfils the following equation of motion,
\begin{eqnarray}
\nabla^2f_R = &-\frac{1}{a}\Omega_m H_0^2\left(\eta - 1\right)  + a^2\Omega_m H_0^2 \times \nonumber\\
&\times \left[\left(1+4\frac{\Omega_\Lambda}{\Omega_m}\right)\left(\frac{f_{R0}}{f_R}\right)^{\frac{1}{n+1}}  - \left(a^{-3} + 4\frac{\Omega_\Lambda}{\Omega_m}\right)\right], 
\end{eqnarray}
where $f_{R0}$ is the value that corresponds to the minimum for the background density today \scb{and $\eta$ is the local matter density in units of the mean density of the Universe.} 

The geodesic equation takes the  form
\begin{equation}
\ddot{\bf{x}} + 2H \dot{\bf{x}} + \frac{\nabla\Phi}{a^2} - \frac{1}{2}\frac{\nabla f_R}{a^2}= 0, 
\end{equation}
where we obtain an extra term with respect to standard gravity which corresponds to the fifth force.

\subsection{The N-body code}

\scb{The simulations were run with the code Isis \citep{2014A&A...562A..78L}, which is a modified gravity version of the code Ramses \citep{2002A&A...385..337T}.  The code is a particle mesh code which includes adaptive mesh refinements and  can thus increase the resolution of the solutions when needed (i.e. in the centre of the dark matter halos).  In order to solve the equations for the scalar field, the code uses a non-linear version of the linear multigrid solver with which Ramses solves the Poisson equation.  The solver works by doing Gauss-Seidel iterations on the discretised version of the equations to find improved solutions based on an initial guess.  Given the multiscale properties of the problem, the solver also makes  iterations on coarse grids, which can increase the speed of convergence by several orders of magnitude.  The code is the only one present in the literature that can solve in an explicit way the scalar field equations out of the static limit.  This non-static solver was already applied to symmetron and disformal gravity models \citep{2014PhRvD..89h4023L, 2015arXiv150407142H}.  A comparison between the accuracy of this code and other codes that solve the same set of equations can be found in \cite{2015arXiv150606384W}. }

\subsection{Simulations}
The data to be used for the analysis was obtained from a set of simulations that were run with both standard gravity and the two modified gravity models.  Table \ref{tab:model_parameters} summarises the model parameters.  The initial conditions were generated assuming that the scalar field is fully screened at high redshift and thus, the Zel'dovich approximation is valid for all the models.  To generate the only set of initial conditions we used the package Cosmics \citep[][]{1995astro.ph..6070B}.  The box size and number of particles employed are $256$ Mpc/h and $512^3$.  The background cosmology is also the same for all the simulations and is defined as a flat $\Lambda$CDM model given by $(\Omega_m, \Omega_{\Lambda}, H_0) = (0.267,0.733,71.9 ~ \text{km/sec/Mpc})$.  The simulations are normalised using the linearly extrapolated value of $\sigma_8$ at redshift zero.  All the simulations have the same normalisation, which is necessary if we want all the simulations to be consistent with the CMB.  The simulations were run up to redshift zero, which is the moment at which we make our analysis. \scb{Furthermore, all the simulations use the same background cosmology with exactly the same initial conditions (i.e.  with the same initial seed for the random numbers generator).  Thus we assume the effects of the scalar field appear only because of the existence of a fifth force in the perturbation level.}

The power spectrum of the initial conditions are not normalised to CMB measurements, but rather are normalised using a redshift zero $\sigma_8$, which is evolved back in time with linear Newtonian theory. This is a standard approach within the community; however, this method has the drawback that the final normalisation of the simulation cannot be completely controlled owing to the presence of the fifth force in the modified gravity models.  In all cases, even changes in clustering amplitudes should be investigated within the domain of modified gravity theories. 

In order to investigate the clustering of dark matter halos, we make use of halo catalogues that were obtained with the phase-space, friends-of-friends (FOF) code {\rm Rockstar} \citep{2013ApJ...762..109B}.  
The samples used for the analysis include all the halos reported by the halo finder with no discrimination between virialized and non-virialized objects.  The halo catalogue has a cut-off for low-mass halos at 20 particles per halo, which corresponds to a minimum halo mass of $1.85\times10^{11}M_{\odot}/h$.  The same cut-off was used for all the simulations. \scb{We have compared the halo mass function from these simulations with higher resolution simulations and with theoretical models and find no bias in using 20 particles to define the halos.}

For details on the implementation of the modified equations on the code Isis and on the simulations we refer the reader to \cite{2014A&A...562A..78L}.

\begin{table}
  \caption{Model parameter values}
  \begin{tabular}{lccc}
   \hline\hline                        
    Model &$\lambda_0$&$z_{SSB}$&$\beta_s$\\
              & {\tiny{Mpc/h}} &  & \\
 \hline    
    Symm A & 1 & 1 & 1\\
    Symm B & 1 & 2 & 1\\
    Symm C & 1 & 1 & 2\\
    Symm D & 1 & 3 & 1\\
     \hline    
\end{tabular}
\begin{tabular}{lccc}
 \hline\hline                        
    Model &$n$ & $|f_{R0}|$ & $\lambda_0$\\
        & & &  {\tiny{Mpc/h}}\\
 \hline
    FOFR4 & 1 & $10^{-4}$& 23.7\\
    FOFR5 & 1 & $10^{-5}$& 7.5\\
    FOFR6 & 1 & $10^{-6}$& 2.4\\
        ~ & ~ & ~ & ~ \\
     \hline    
  \end{tabular}
  \label{tab:model_parameters}
\end{table}

\section{Correlation statistics}\label{stats}
We want to quantify the spatial clustering of the simulated halos so that comparisons  between models are easier to perform. To achieve this we calculate 
 the two- and three-point correlation functions using well-known estimators, which are described below.

\subsection{Two-point correlation function}
We estimate the full two-point correlation function (2PCF)  of the DM halo distribution in each of the simulations listed in Table \ref{tab:model_parameters}. The 2PCF is estimated in real and redshift space and in both isotropic and anisotropic $\sigma,\pi$-decompositions. 
The correlation functions are calculated using the ``Landy-Szalay'' estimator \citep{1993ApJ...412...64L}, 
\begin{equation}
\xi(\vec{r})=\frac{DD(\vec{r})-2DR(\vec{r})+RR(\vec{r})}{RR(\vec{r})},
\label{eq:2pcf}
\end{equation}
where $DD(\vec{r})$ is the number of data--data pairs, $DR$ the number of data-random pairs, and $RR$ is the number of random--random pairs, all  separated by a displacement vector $\vec{r}$ and properly normalised. In the isotropic case we measure $\xi(r)$ with no angular dependence and we choose the binning $r=0\rightarrow60$Mpc/h in 20 linearly spaced bins. \scb{The number of random particles used to define the unclustered reference sample is $\approx$20 times the number of halo particles. This is done to reduce statistical fluctuation due to Poisson noise in the random pair counting.}

In the anisotropic analysis we decompose the vector $\vec{r}$ into a line-of-sight distance component $\sigma$ and a perpendicular component $\pi$. Then we proceed to measure $\xi(\sigma,\pi)$ from $\sigma$ and $\pi=0\rightarrow60$ Mpc in 15 linearly spaced bins, resulting in 225 bins in the $\sigma-\pi$ space. 

\subsection{Three-point correlation function}
Higher-order correlations are usually denoted  $\xi_{n}(r_{1},....,r_{n})$, where $n$~is the order of the correlation function. As an example, the 3PCF is defined as the joint probability of there being a galaxy in each of the volume elements $dV_{1}$, $dV_{2}$, and $dV_{3}$ given that these elements are arranged in a configuration defined by the sides of the triangle, ${\bf r}_{1}$, ${\bf r}_{2}$,  and ${\bf r}_{3}$. The joint probability can be written as
\begin{equation}
dP_{1,2,3}=\bar{n}^{3}[1+\xi({\bf r}_{1})+\xi({\bf r}_{2})+\xi({\bf r}_{3})+\zeta({\bf r}_{1},{\bf r}_{2},{\bf r}_{3})]dV_{1}dV_{2}dV_{3}.
\label{eq:3PCF}
\end{equation}
The expression above consists of several parts:  the sum of correlations for each side of the triangle and $\zeta$, the full three-point correlation function, and $\bar{n}$  the mean density of data points.  We utilise the 3PCF estimator of \citet{1998ApJ...494L..41S}, 
\begin{equation}
\zeta = \frac{DDD - 3DDR + 3DRR - RRR}{RRR},
\label{eq:salay}
\end{equation}
where each term represents the normalised triplet counts in the 
data (D) and random (R) fields that satisfy a particular 
triangular configuration.

The 3PCF is computed in an isotropic (angle-averaged) way and is a function of three variables that uniquely define a triangular configuration. The shape parameters can either be the three sides of the triangle, $(r_1,r_2,r_3)$, or more commonly $(s,q,\theta), $ where
\begin{eqnarray}
s&=&r_1,\\
q&=&\frac{r_2}{r_1},\label{eq:q}\\
\theta&=&\cos^{-1}(\hat{r}_1.\hat{r}_2) \label{eq:theta}.
\end{eqnarray}
In Eq~\ref{eq:theta}, $\hat{r}_{i}$ is the unit vector of side $i$ of the triangle. The 3PCF is usually calculated in various configurations where  $s$~and $q$~are fixed, while $\theta$ is varied from $0^\circ$~to $180^\circ$.

In our analysis we focus on two triangular configurations, $s,q=2{\rm Mpc}/h,1$ and $s,q=3{\rm Mpc}/h$,2 with each probing eight equally spaced bins in $\cos{\theta}$. Our choice of scales was determined by practical and statistical issues. On small scales, baryonic physics can be significant; however, our simulations do not include these effects, so comparison to observation on scales less than 1Mpc would be problematic. On larger scales the 3PCF amplitude drops off drastically and owing to the small box size of our simulation, the statistical noise becomes significant. We find through testing that s=2,3Mpc and q=1,2 give stable results.

\section{Results}\label{results}
\begin{figure*}
\centering
\includegraphics[width=0.9\columnwidth]{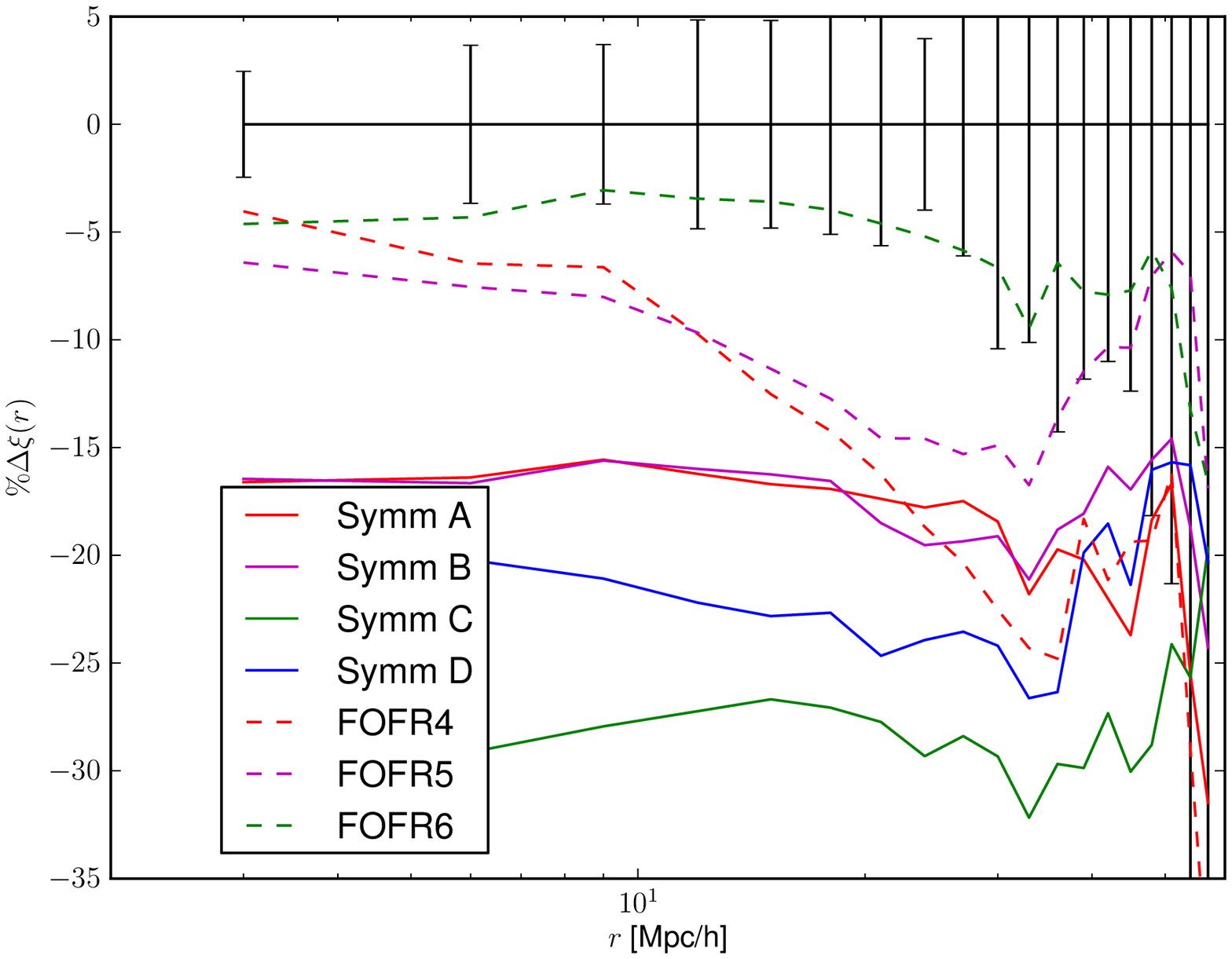}
\includegraphics[width=0.9\columnwidth]{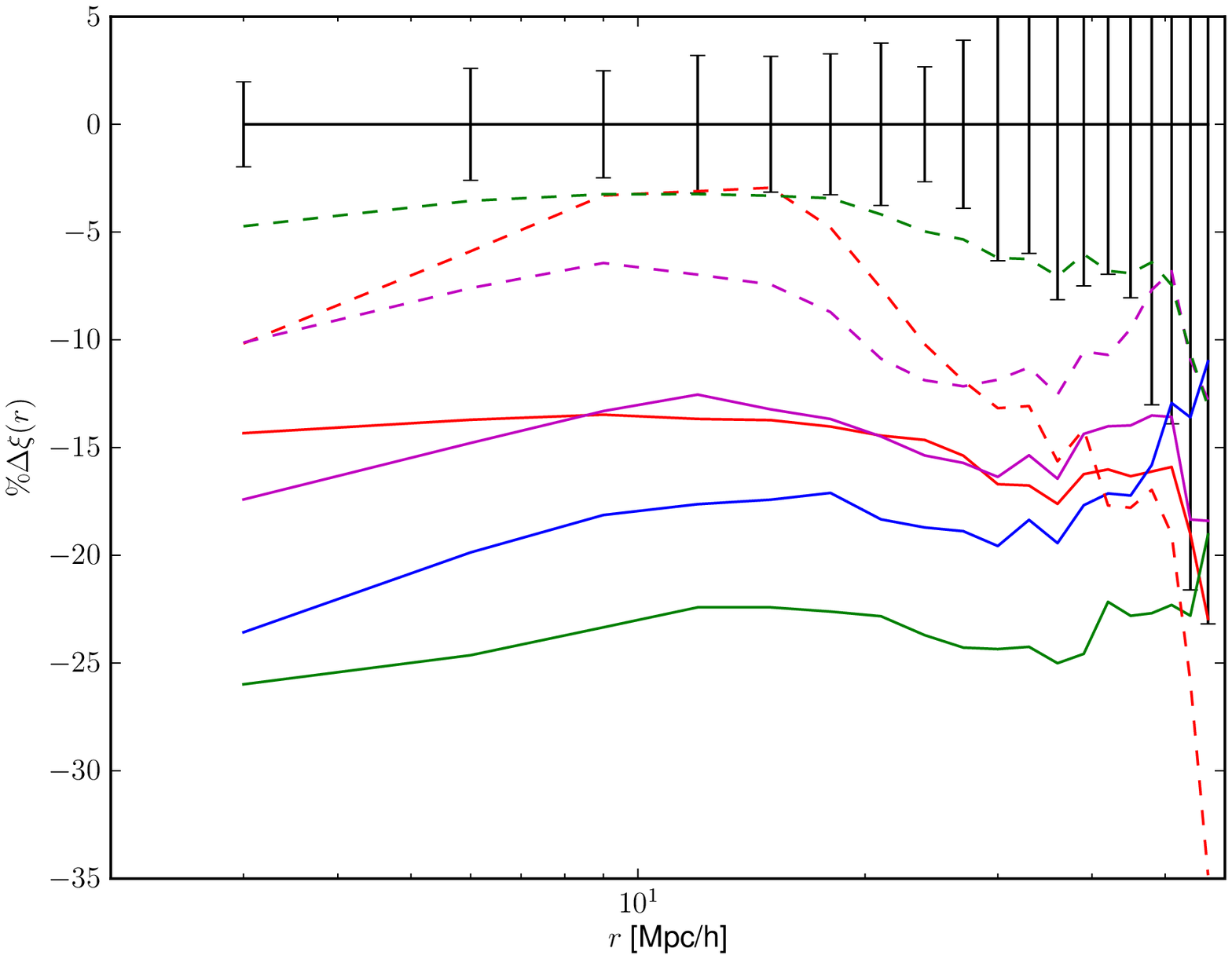}
\caption{\label{fig:2pcf}{\em left:} Fractional difference between the 2PCF measurements of dark matter halos within modified gravity and $\Lambda$CDM simulations in real comoving space. {\em right:} same as left plot, now in redshift space. }
\end{figure*}

In this section we will present the results of our clustering statistics for the various models, in real and redshift spaces. 

\subsection{Two-point function results}
\scb{In Fig.\ref{fig:2pcf} we show the relative difference $\Delta\xi=(\xi^{mod}-\xi^{\Lambda CDM})/\xi^{\Lambda CDM}$ between the 2PCF (calculated from Eq.~\ref{eq:2pcf}) of the modified gravity simulations (see  \S\ref{sims}) and the $\Lambda$CDM case} . The left-hand plot shows the relative difference between the 2PCF of the dark matter halos in real comoving space, while the right-hand plot shows the 2PCF in redshift space. 

The errors on $\xi(r)$ were derived using the jackknife method~\cite{2002ApJ...579...48S}, which involves dividing the simulation box into $N$ subsections with equal volume and then computing the mean and variance of $\xi(r)$ from these $N$ measurements of the correlation function with the $i^{th}$ region removed each time (where $i=1...N$). In our analysis, we choose $N=27$ and determine the  variance from~\cite{1993stp..book.....L}, 
\begin{equation}
\sigma^2_\xi(r_i)=\frac{N_{jk}-1}{N_{jk}}\sum^{N_{jk}}_{k=1}[\xi_k(r_i)-\overline{\xi}(r_i)]^2,
\end{equation}
where $N_{jk}$ is the number of jackknife samples used and $r_i$ represents a single bin in the $\xi(r)$ configuration space and $\overline{\xi}(r_i)$ is the mean of the jackknifed samples. Although not exact, the jackknife variance estimate will inform us if any deviations in the statistics between the $\Lambda$CDM model and modified gravity are significant and above the sampling noise. Of course, this variance will not inform us of the detectability of any deviation in a particular current or future observational dataset.

Considering first the real-space 2PCF in the left plot of Fig.\ref{fig:2pcf} all the models, except for the weakest f(R) model FOFR6, show significant deviation from the $\Lambda$CDM case on scales $<30$ Mpc/$h$. In redshift space, we observe a similar amount of deviation from the $\Lambda$CDM case, with slightly more suppression on small scales.

The shapes of these curves suggest two things:  The small-scale clustering amplitude of the modified gravity simulations may be slightly lower than that of $\Lambda$CDM, and  the non-linear small-scale velocities may be higher in the modified gravity simulations, which would suppress clustering on small scales.
\scb{Our results in Fig.\ref{fig:2pcf} for the FOFR4 model appear consistent with the level of decrement observed in \citet{2014PhRvD..90l3515T} where they find $\Delta\xi\approx13\%$ on scales $\sim$50 Mpc/$h$}. {The lower amplitude in both the real and redshift-space 2PCF is mainly due to the bias in the modified gravity simulation being lower than $\Lambda$CDM. This behaviour of the bias in modified gravity has already been suggested in the literature, where a decrease relative to $\Lambda$CDM in the dark matter halo bias in f(R) gravity was pointed out by \citet{2009PhRvD..79h3518S} and likewise for the galaxy bias by \citet{2013MNRAS.436.2672F}. However, to fully understand this issue is beyond the scope of this paper; we thus focus our attention mainly on the differences  between real and redshift-space quantities}

\begin{figure}
\centering
\includegraphics[width=0.9\columnwidth]{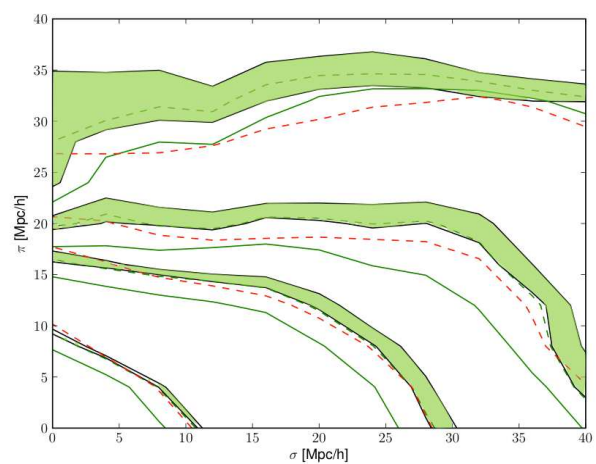}
\caption{\label{fig:sigmapi} Anisotropic, redshift-space, 2PCF measurements for three modified gravity simulations, using the dark matter halos in redshift space. For clarity we show only two extreme cases, FOFR4 (red dashed line) and SymmC (green solid line), and a mildly deviating model FOFR6 (light green dashed line).
The four contour groups correspond to 2PCF values of [0.4, 0.1, 0.05, 0.01]. The green shaded region denotes the 1$\sigma$ error bound around the  $\Lambda$CDM value.}
\end{figure}

So far we have looked at the isotropic 2PCF, however, this is not the most sensitive statistic to the redshift-space distortion effects. In Fig.\ref{fig:sigmapi} we show the anisotropic 2PCF in redshift space, decomposed into distances $(\sigma,\pi),$ which correspond to directions across and along the line of sight. 

The four contours correspond to the values [0.4, 0.1, 0.05, 0.01]. The green shaded region denotes the 1$\sigma$ error bound around the  $\Lambda$CDM value. For clarity we only show the clustering contours for three modified gravity models, two that exhibit the most deviation from $\Lambda$CDM (FOFR4 and SYMMC) and one that shows very similar clustering signal to $\Lambda$CDM (FOFR6). The various modified gravity simulations show very similar contours in this projection, although contours lying close to the line of sight exhibit some dispersion, i.e. a stronger fingers-of-god (FoG) effect compared to $\Lambda$CDM. Once again the \rm{FOFR4} model shows significant deviation from $\Lambda$CDM, especially on scales below 30 Mpc$/h$. {However, it is the SYMMC model that shows the most dramatic deviation between the models investigated and  $\Lambda$CDM}.  This effect means that owing to the presence of a fifth force, velocities are in general lager in modified gravity theories \citep{2015MNRAS.449.2837G, 2014PhRvL.112v1102H}.  We note that redshift space distortions were already studied in $f(R)$ gravity by \citet{2012MNRAS.425.2128J}; however, that work focuses its attention on the power spectrum of dark matter density perturbations.

\subsection{Three-point function results}
\begin{figure*}
\centering
\includegraphics[width=0.95\columnwidth]{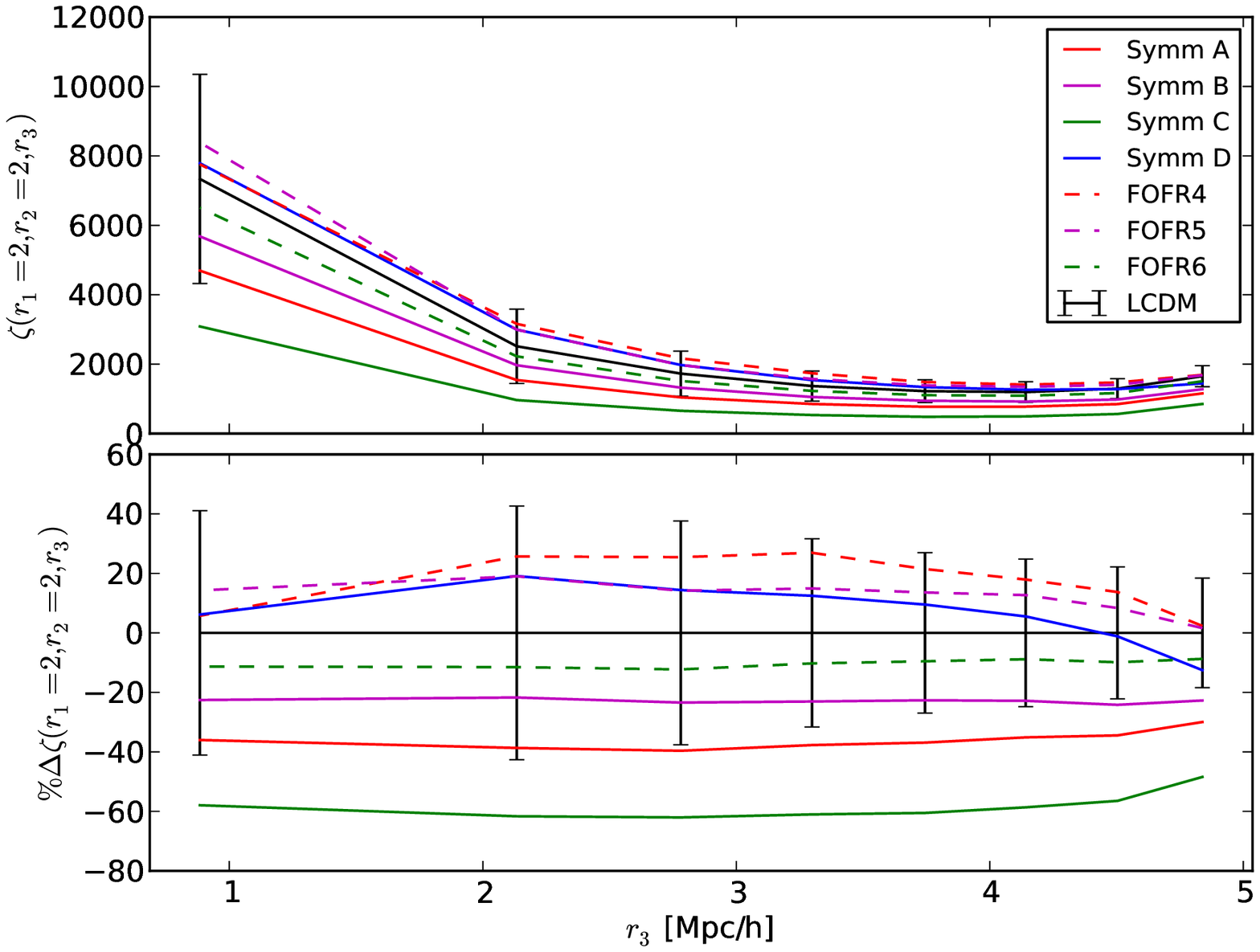}
\includegraphics[width=0.95\columnwidth]{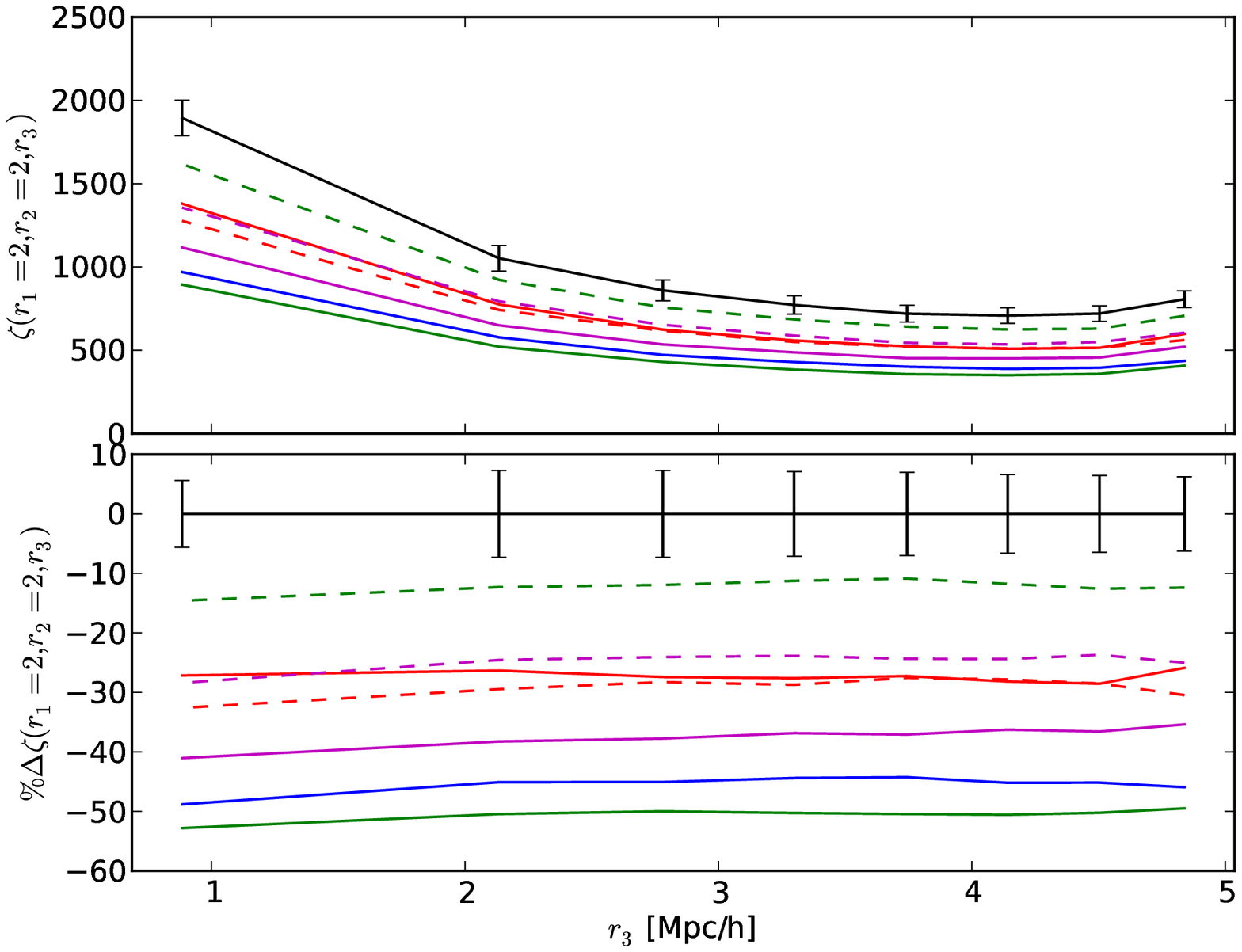}
\caption{\label{fig:2_2}{\em left:}  Halo 3PCF for the eight simulations using the triangular configuration of s=$2 \rm{Mpc}/h$, q=1, in real space. The full 3PCF is shown in the top panel, while the relative difference compared to  $\Lambda$CDM is in the lower panel. {\em  right:}  Same as right-hand panel, but in redshift space. }
\end{figure*}

In the upper panels of Fig.\ref{fig:2_2} we show the 3PCF for all the simulations in the triangular configuration with $s=2 ~\mathrm{Mpc}/h$ and $q=1$, in real space (left panel) and redshift space (right panel). The  length of the sides for this configuration are in the ranges $1.5<r_1<2.5, 1.5<r_2<2.5, 0<r_3<5$.

The lower panels of Fig.\ref{fig:2_2} show the relative difference, $\Delta\zeta$, between the 3PCF of $\Lambda$CDM and modified gravity, where $\Delta\zeta= (\zeta^{mod} - \zeta^{\Lambda CDM}) / \zeta^{\Lambda CDM}$. The errorbars are once again estimated via 27 jackknife samples, and are basically determined by the comic variance of the structures within our 256 Mpc/h size simulation box.  {Nevertheless, smaller deviations between models can be significant in a comparison between simulations since all runs use the same phases and amplitudes in the initial conditions.}

We can compare these results to Fig. 9 of \citet{2005ApJ...632...29F} who look at the same triangular configuration.  Our results differ by  a factor of $\sim$4-8 because  we  consider the halo correlations, while in \citet{2005ApJ...632...29F}  they consider the dark matter clustering, and also because the precise triangular binning scheme may differ between these two analyses.

We observe a dispersion between models which suggests that the 3PCF alone is at least as powerful a probe of modified gravitational clustering as the 2PCF. However, considering that the 3PCF is sensitive to the bias in a different way than the 2PCF, there may also be the potential to break any degeneracy between bias and modified gravity by using a combination of the 2PCF and 3PCF.

In real space the 3PCF of this triangular configuration shows mild deviation from the $\Lambda$CDM model. However, in redshift space, the 3PCF of this configuration shows a larger variation between models and all the modified gravity models deviate from $\Lambda$CDM; this is particularly evident in the FOFR models and SYMMD.

\begin{figure*}
\centering
\includegraphics[width=0.95\columnwidth]{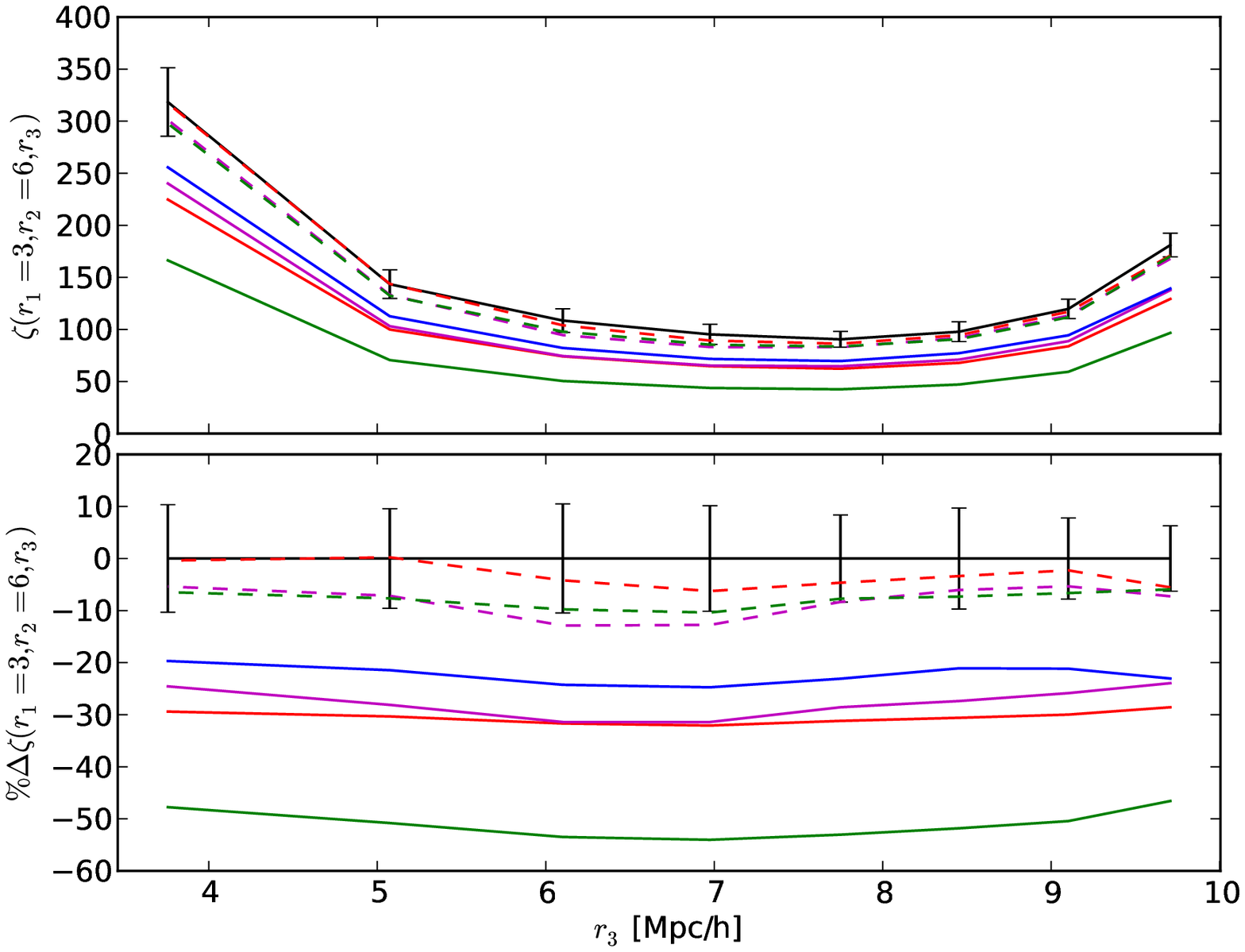}
\includegraphics[width=0.95\columnwidth]{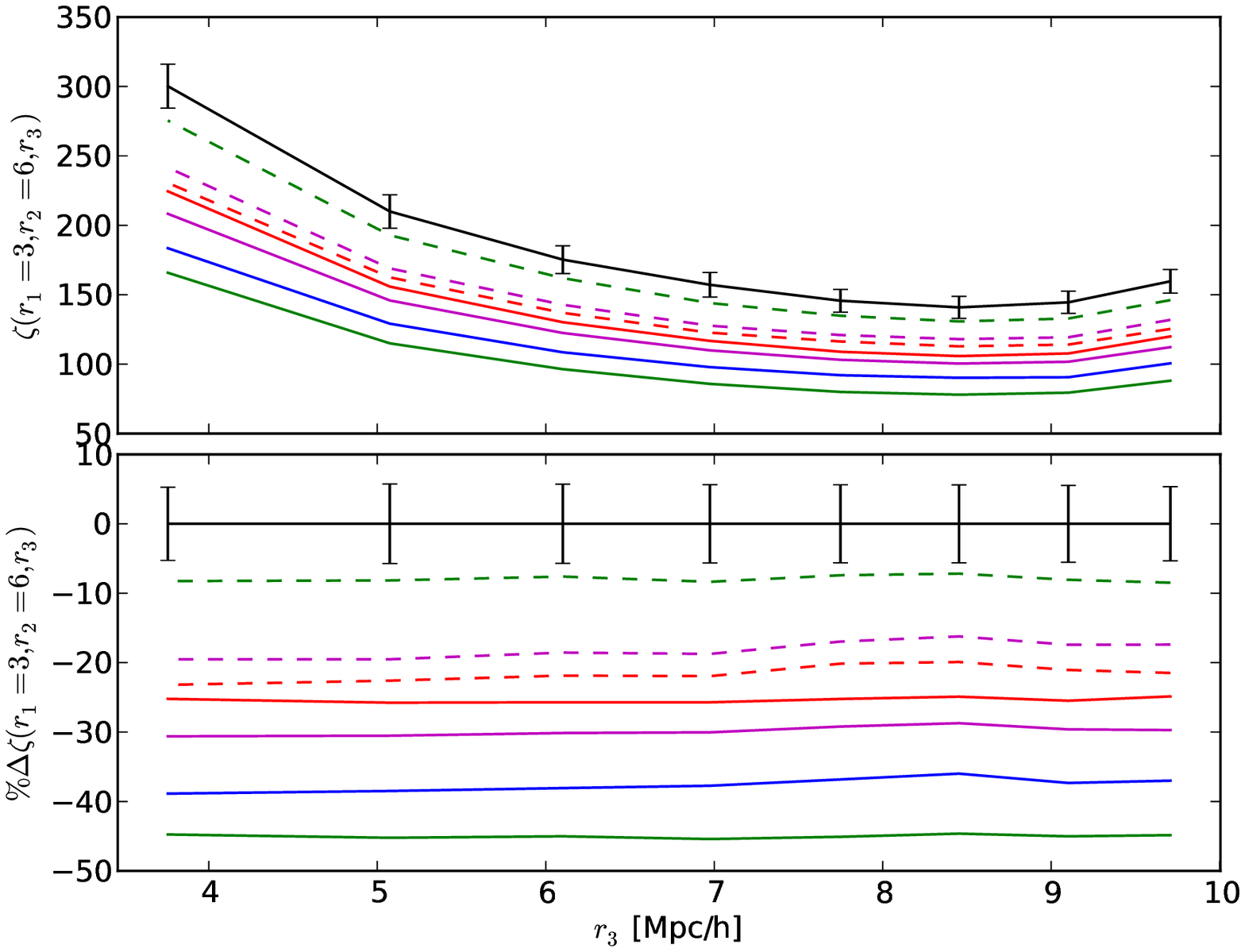}
\caption{\label{fig:3_6} Same as Fig.\ref{fig:2_2} with triangular configuration with s=$3 \rm{Mpc}/h$, q=2 }
\end{figure*}

In Fig.\ref{fig:3_6} we investigate the 3PCF with triangular configuration $s=3~ \mathrm{Mpc}/h$ and $q=2$. In this case the side lengths correspond to $2.5<r_1<3.5, 5.5<r_2<6.5$, and $2<r_3<10$ and thus this configuration will preferentially select triplets that occupy three distinct halos. However, in redshift space the FoG effect can bring in a one-halo contribution for ``closed'' and ``open'' triangles (i.e. when $r_3$ is 3 and 9 Mpc$/h$, respectively), although the latter may be on too large a scale to add any significant effect. The one-halo contribution, within the halo model framework \citep{2002PhR...372....1C}, is the clustering signal in the 2PCF (or 3PCF) of pairs (or triplets) of galaxies residing in the same DM halo. This is opposed to the other contributions from galaxies residing in separate halos. In redshift space the clusters of galaxies can be elongated along the line of sight, thus the one-halo term can be significant on slightly larger scales than it would be in real-space.

In the left-hand panels of Fig.\ref{fig:3_6} we see that in real space the symmetron models are already significantly deviated from $\Lambda$CDM, while the FOFR models all lie within  1-$\sigma$ of  $\Lambda$CDM. In redshift space, shown in the right panel, the models show a similar behaviour; however, the FOFR models have now diverged from  $\Lambda$CDM. Interestingly the symmetron models A, B, and C do not show much relative difference between real and redshift space in contrast to SYMMD which shows a significant shift. 

Our real space clustering measurements support the conclusion of \cite{2011JCAP...11..019G}, where they find a weak dependence of the real space reduced bispectrum on $|f_{R0}|$. However in \cite{2011JCAP...11..019G} they look at scales  outside the cluster size, i.e.  $k\approx0.4$. In our work looking at the real and redshift-space full 3PCF on small scales, we find significant departures from the LCDM clustering. We suspect that this is due to the altered velocity distribution in modified gravity which results in anisotropic redshift space distortions that can be detected with higher-order clustering statistics. However, this claim should be tested further in future work.

\section{Conclusions}\label{conclusions}
We present the first measurements of the higher-order clustering in redshift space of certain modified gravity models. Using the three-point correlation function we are able to probe deviations in the clustering signal between modified gravity models and $\Lambda$CDM.

Owing to the volume and number of numerical simulations that we used, we show only a qualitative study of the higher-order clustering in redshift space. We have shown that the small-scale clustering in redshift space is significantly altered in the modified gravity models we consider  compared to $\Lambda$CDM. This is especially true for higher-order statistics, namely the 3PCF. 

In this work we have assumed a single set of cosmological parameters while varying the gravitational force; however, the effect of the modified gravity on the two- and three-point clustering should be quantified precisely and checked for any degeneracy with the cosmological parameters. This will require many N-body simulations and was therefore beyond  the scope of our study; however, we hope that this work will prompt further research in this direction.

\subsection*{Acknowledgments}
Several of the authors thank Arman Shafieloo, Stephen Appleby, Dhiraj Hazra, and the Asia Pacific Center for Theoretical Physics for their hospitality during their 1st joint workshop on Dark Energy, where this project was realised. We thank Prof. Juhan Kim for advising us on the halo mass function in our N-body simulations. We also thank the anonymous referee for very helpful comments and suggestions.
DFM  acknowledges support from the Research Council of Norway. CLL acknowledges support from the Research Council of Norway through grant 216756 and from STFC consolidated grant ST/L00075X/1. Numerical calculations were performed using a high-performance computing cluster at the Korea Institute for Advanced Study (KIAS Center for Advanced Computation Linux Cluster System), and the cluster Hexagon, which is part of the NOTUR facilities in Norway. 

\bibliographystyle{aa}
\bibliography{references}

\end{document}